\documentclass[letterpaper, 10 pt, conference]{ieeetran}  

\IEEEoverridecommandlockouts                              
\usepackage{times}
\usepackage{graphicx,psfrag}
\usepackage{amsmath,amssymb,amsfonts,mathrsfs}
\usepackage{color,epstopdf}
\usepackage{algorithmic}

\def\qedp{\hspace*{\fill}~{\tiny $\blacksquare$}}

\newtheorem{remark}{Remark}
\newtheorem{assumption}{Assumption}

\newtheorem{theorem}{Theorem}

\newtheorem{proposition}{Proposition}

\newcommand\myrule{\hrule width \columnwidth height .4pt}

\newcommand{\until}[1]{\{1,\dots, #1\}}
\newcommand{\subscr}[2]{#1_{\textup{#2}}}

\newcommand{\setdef}[2]{\{#1 \; : \; #2\}}

\newcommand{\map}[3]{#1: #2 \rightarrow #3}

\newcommand{\integernonnegative}{\ensuremath{\mathbb{Z}}_{\ge 0}}
\newcommand{\real}{\ensuremath{\mathbb{R}}}
\newcommand{\realpositive}{\ensuremath{\real_{>0}}}
\newcommand{\realnonnegative}{\ensuremath{\real}_{\geq0}}

\newcommand{\R}{\mathbb{R}}

\newcommand{\eps}{\varepsilon}
\newcommand{\E}{\mathcal{E}}    

\newcommand{\neigh}[1]{{\cal N}_{#1}}
\renewcommand{\S}{\mathcal{S}}    

\newcommand{\degmax}{\subscr{d}{max}}

\renewcommand{\deg}{d}

\renewcommand{\S}{{ \mathcal{S} \stopmodif}}
\newcommand{\ave}{\operatorname{ave}} %
\newcommand{\sign}{\operatorname{sign}} %

\newcommand{\dst}{\displaystyle}

\def\be{\begin{equation}}
\def\ee{\end{equation}}
\def\ba{\begin{array}}
\def\ea{\end{array}}
\def\eqa{\begin{eqnarray}}
\def\eqe{\end{eqnarray}}

\def\stopmodif{\color{black}}

\title{
Self-triggered Coordination over a Shared Network under Denial-of-Service
}

\author{Danial Senejohnny, Pietro Tesi, Claudio De Persis
\thanks{D. Senejohnny, P. Tesi, and C. De Persis are with ENTEG, Faculty of Mathematics and Natural
Sciences, University of Groningen, 9747 AG Groningen, The Netherlands
{\tt\small d.senejohnny@rug.nl, p.tesi@rug.nl, c.de.persis@rug.nl}
}
}

\begin{document}

\maketitle
\thispagestyle{empty}
\pagestyle{empty}

\begin{abstract}

The issue of security has become ever more prevalent in the analysis and design of cyber-physical systems. In this paper, we analyze a consensus network in the presence of Denial-of-Service (DoS) attacks, namely attacks that prevent communication among the network agents. 
By introducing a notion of Persistency-of-Communication (PoC), we provide a characterization of DoS frequency and duration such that consensus is not destroyed.
An example is given to substantiate the analysis.

\end{abstract}

\section{Introduction}

In recent years, the issue of security has become ever more prevalent in the analysis and design of cyber-physical  systems (CPSs), namely systems that exhibit a tight conjoining of computational resources and physical resources.
As argued in \cite{Sastry-survey, Sinopoli}, security in CPSs drastically differs from security in general-purpose computing systems since attacks can cause disruptions that transcend the cyber realm and affect the physical world. In CPSs, attacks to the communication links can be classified as either deception attacks or Denial-of-Service (DoS) attacks. The former affect the trustworthiness of data by manipulating the packets transmitted over the network; see \cite{Tabuada}-\cite{ICM-Geometric} and the references therein. DoS attacks are instead primarily intended to affect the timeliness of the information exchange, \emph{i.e.}, to cause packet losses; see for instance \cite{DoS-1,DoS-2} for an introduction to the topic. This paper is concerned with DoS attacks. 

In the literature, the problem of securing robustness of CPSs against DoS has been investigated by many research groups \cite{Andre}-\cite{Martinez}. In these papers, however, the analysis is restricted to a \emph{centralized} setting, namely to a classical (single-loop) plant-controller configuration. On the other hand, no quantitative results are available for \emph{distributed} settings. The purpose of this paper is to explore this topic.

We investigate the issues of DoS with respect to \emph{consensus} networks. Specifically, inspired by \cite{Claudio-Paolo}, we consider a \emph{self-triggered} consensus network. At each sampling time, a certain subset of active agents poll their neighbors obtaining relative measurements of the consensus variable of interest: the available information is then used by the active agents to update their controls and compute their next update times. 
The attacker objective is to prevent consensus by denying communication among the agents. 
Consensus is a prototypical problem in distributed settings with a huge range of applications, spanning from formation and cooperative robotics to surveillance and distributed computing; see for instance \cite{Claudio-Paolo}-\cite{Cortes}. 
On the other hand, self-triggered coordination turns out to be of major interest when consensus has to be achieved
in spite of possibly severe communication constraints. In this respect, a remarkable feature of self-triggered 
coordination lies in the possibility of ensuring consensus properties in the absence of any
global information on the graph topology and with no need to synchronize the agentsÕ local clocks.

A basic question in the analysis of distributed coordination in the presence of DoS is concerned with the modeling of DoS attacks. In this paper, no assumption is made regarding the DoS attack underlying strategy. We consider a general attack model that only constrains the attacker action in time by posing limitations on the frequency of DoS attacks and their duration. This makes it possible to capture many different types of DoS attacks, including trivial, periodic, random and protocol-aware jamming attacks \cite{A}-\cite{D}. 
One contribution of this paper is an explicit characterization of the frequency and duration of DoS attacks under which consensus properties can be preserved. The result is intuitive as it relates consensus with the ratio between the on/off periods of jamming. 

The analysis taken here reminds of stability problems for switching systems. More specifically, since DoS generates communications failures, the problem is naturally casted as a consensus problem for switching topology networks. This problem is certainly not new in the literature. For instance, \cite{Murry} shows that agreement can be reached whenever the graph connectivity is preserved point-wise in time; \cite{Arcak} suggests a \emph{Persistency of Excitation} (PoE) condition, which stipulates that graph connectivity be established over a period of time, rather than point-wise in time, which is similar to joint connectivity assumption in \cite{Morse}. 
In CPSs, however, the situation turns out to be drastically different. In fact, one needs to deal with the fact that networked communication is inherently digital, which means that the rate at which the transmissions are scheduled cannot be arbitrarily large. 
Under such circumstances, the aforementioned tools turn out be ineffective. 

In order to cope with this situation, we introduce a notion of \emph{Persistency of Communication} (PoC). This notion naturally extends the PoE condition to a digital networked setting by requiring that the graph connectivity
be established over periods of time that are consistent with the constraints imposed by the communication medium.
A characterization of DoS frequency and duration under which consensus properties can be preserved is then obtained  by exploiting the PoC condition. 

The remainder of this paper is organized as follows. In
Section II we  formulate the control problem along with a characterization of the considered class of DoS signals.  The main results in Section III. A numerical example is given in  Section IV. Finally, Section V ends the paper with concluding remarks.

\emph{Notation}. The notation adopted in this paper is in the main standard.
We denote by $\real$, $\realpositive$, $\realnonnegative$ the sets of real, positive, and nonnegative numbers, respectively. Also, we denote by $\integernonnegative$ the set of nonnegative integers.

\section{Problem Formulation}\label{sec:PF}

\subsection{Distributed Control System} 

We assume to have a set of nodes $I=\until{n}$ representing our agents and an undirected connected \stopmodif  graph $G=(I,E)$ with $E$ a set of unordered pairs of nodes, called edges. We denote by $D$ and $L$ the Incidence and Laplacian matrix of $G$, respectively, where the latter is a symmetric matrix. For each node $i\in I$, we denote by $\neigh{i}$ the set of its neighbors, and by $\deg_i$ its degree, that is, the cardinality of $\neigh{i}.$

We consider the following hybrid dynamics on a triplet of $n$-dimensional variables involving the consensus variable $x$, the controls $u$, 
and the local clock variables 
$\theta$. All these variables are defined for time $t\ge0$.
Controls are assumed to belong to $\{-1,0,+1\}$. The specific quantizer of choice is $\map{\sign_\eps}{\real}{\{-1,0,+1\}}$, defined according to 
\be\label{eq:signeps} \sign_\eps(z)=\begin{cases} \sign(z) & \text{if}\: |z|\ge\eps\\
0 & \text{otherwise}
\end{cases}\ee
where $\eps>0$ is a sensitivity parameter, which can be used at the design 
stage for trading-off frequency of the control updates vs. accuracy of the 
consensus region.

The system $(x,u,\theta)\in\real^{3n}$ in the \emph{nominal operating
mode}, \emph{i.e.}, in the absence of DoS,
satisfies the following continuous evolution
\be\label{eq:modelloA-cont} \begin{cases}
\dot x_i=u_i\\
\dot u_i=0\\
\dot \theta_i=-1
\end{cases}\ee
except for every $t$ such that the set 
$\S(\theta,t)=\setdef{i\in I}{\theta_i=0}$ 
is non-empty. At such time instants, the system satisfies the following discrete evolution
\be\label{eq:modelloA-discAA}
\begin{cases} x_i(t^+)=x_i(t) \quad \forall i\in I \\
u_i(t^+)=
\begin{cases}
\sign_{\eps}\!\left(\ave_i(t)\right) \quad \text{if}\: i\in \S(\theta,t)\\
u_i(t) \quad \text{otherwise}
\end{cases}\\
 \theta_i(t^+)=\begin{cases} f_i(x(t)) \quad \text{if}\: i\in \S(\theta,t)\\
\theta_i(t) \quad \text{otherwise}
\end{cases}\\
\end{cases}
\ee
where for every $i\in I$ the map $\map{f_i}{\real^n}{\realpositive}$ is defined by

\be \label{eq:sam}
f_i(x(t))=\left\{\ba{ccc}
\dst\frac{|\ave_i(t)|}{4d_{i}} & {\rm if}  &
|\ave_i(t)|\ge \eps \\ [4mm]
\dst\frac{\eps}{4d_{i}} 
& {\rm if} & |\ave_i(t)|< \eps \ea\right.,
\ee
where, for conciseness, we have defined
\be\label{avei}
\ave_i(t)=\sum_{j\in\neigh{i}}(x_j(t)-x_i(t)).
\ee

Self-triggered coordination algorithms such as 
(\ref{eq:modelloA-cont})-(\ref{eq:sam}).
turn out to be of major interest when consensus has to be achieved in spite of possibly severe communication constraints. In this respect, a remarkable feature of self-triggered coordination lies in the possibility of ensuring consensus properties in the absence of any global information on the graph topology and with no need to synchronize the agents local clocks \cite{Claudio-Paolo}.

The result which follows characterizes the convergence 
properties of (\ref{eq:modelloA-cont})-(\ref{eq:sam}) in the nominal 
operating mode, and will serve as a basis for the 
developments of the paper.

\begin{theorem} 
\label{thm:modelloA-convergenceAA}
\cite{Claudio-Paolo}. Given any $\bar x \in \mathbb R^n$, let $x(t)$ be the solution to~\eqref{eq:modelloA-cont}-\eqref{eq:sam} with $x(0)=\bar x$. 
Then $x(t)$ converges in finite time to a point $x^* \in \mathbb R^n$ belonging to the 
set 
\be\label{set.E}
\E=
\setdef{x\in \R^n}{|\sum_{j\in \neigh{i}} (x_j-x_i)|< \eps \,\, \forall\,i\in I}.
\ee
\qedp
\end{theorem}


\subsection{Denial-of-Service}
We shall refer to DoS as the phenomenon by which 
communication across the network is not possible.
More specifically, we assume that the network nodes make use of
a shared communication medium. Under DoS, none of 
network nodes can send or receive information.  
This scenario is representative of several possible DoS 
threats.  In order to maintain continuity, a discussion on 
this point is deferred to Section II-C. 
Here, we proceed with the DoS modeling and introduce
a number of assumption on its frequency and duration.

Let $\{h_n\}_{n\in \integernonnegative}$, where ${h_0} {\ge 0} $, denote the sequence of DoS off/on transitions, \emph{i.e.}, the time instants at which DoS exhibits a transition from zero (communication is possible) to one (communication is interrupted). Then
\begin{equation} \label{eq:Hn}
H_n: = \{h_n\} \cup \left[ h_n,h_n + \tau _n \right[ 
\end{equation}
represents the $n$-th DoS time-interval, of a length $\tau_n\in {\realpositive}$, over which communication is not possible. Here and in the sequel, it is 
understood that $h_{h+1}>h_n + \tau _n$ for all $n\in \integernonnegative$,
otherwise $H_n \cup H_{n+1}$ could be regarded as a single 
DoS interval.

Given ${t,\tau}\in\real{ \ge 0}$, with $t \ge \tau$, let 
\begin{equation} \label{eq:Xi}
\Xi (\tau ,t) :=\mathop  \bigcup \limits_{n\in{\integernonnegative}} H_n \bigcap {[\tau,t ]}
\end{equation}
represent the sets of time instants where communication is denied and 
\begin{equation}\label{eq:Theta}
\Theta (\tau ,t) := [\tau,t ] \;  \backslash \; \Xi (\tau ,t) 
\end{equation}
represent the sets of time instants where communication is allowed, where $\backslash$ denote the relative complement and accounts for the set of all elements belonging to the time interval $[\tau,t [$, but not to the set $\Xi (\tau ,t)$.

In connection with the definition of the DoS sequence in (\ref{eq:Hn}),
the first question to be addressed is that of determining 
the amount of DoS that the network can tolerate 
before consensus, as defined in Theorem 1, is lost. In this respect, 
it is simple to see that such an amount is 
not arbitrary, and that suitable conditions 
must be imposed on both DoS frequency and duration. 

Let us first consider the frequency at which DoS can occur. 
First notice that $\varepsilon/4d_i$ provides a lower bound on the 
inter-sampling rate of the $i$-th node of the network,
as imposed by the communication medium.
Let now
$\Lambda_n=h_{n+1}-h_n$, with $n\in \integernonnegative$, 
denote the time elapsing between any two successive DoS triggering. 
By letting $d_{min} = \min_{i \in I}d_i$, one immediately sees that if 
\[
\Lambda_n \le \Delta_* := \frac{\varepsilon}{4d_{min}}
\]
then consensus could be destroyed irrespective of the 
adopted communication strategy. This is because DoS would be allowed 
to occur at a rate faster than or equal to the sampling rate 
of some network node, which would clearly preclude the 
possibility to achieve consensus. 
It is intuitively clear that, in order to get stability, the frequency at which DoS can occur must be sufficiently small compared to sampling rate of the network nodes. 
A natural way to express this requirement is via the concept of average dwell-time, as introduced by \cite{Hispana-Morse}. Given ${t,\tau}\in\real{ \ge 0}$ with $t \ge \tau$, let $n(\tau ,t)$ denote the number of DoS off/on transitions occurring on the interval $[\tau,t [$.\

\begin{assumption}[DoS frequency]\label{asum:DoSf}
There exist $\eta  \in\real_{ \ge 0}$ and ${\tau _f} \in\real_{ > {\Delta_*}}$ such that
\begin{equation}\label{eq:f}
n(\tau ,t) \le \eta  + \frac{t - \tau }{\tau _f}
\end{equation}
for all ${t,\tau}\in\real_{ \ge 0}$ and $t \ge \tau$. \qedp
\end{assumption}

In addition to the DoS frequency, one also need to enforce constraints on the DoS duration, namely the length of the intervals over which communication is interrupted. To see this, consider for example a DoS sequence consisting of the singleton 
$\{h_0\}$. Assumption 1 is clearly satisfied with $\eta \geq 1$. However, if 
$H_0=\mathbb R_{\geq 0}$
(communication is never possible) then stability is lost regardless of the adopted control update policy. Recalling the definition of the set $\Xi$ in 
(\ref{eq:Xi}), the assumption that follows provides 
a quite natural counterpart of Assumption 1 with respect to the DoS duration.

\begin{assumption}[DoS Duration]\label{asum:DoSd} 
There exist $\kappa  \in\real_{ \ge 0}$ and ${\tau _d} \in\real_{ > 1}$such that
\begin{equation} \label{eq:d}
\lvert \Xi (\tau ,t) \rvert  \le \kappa  + \frac{t - \tau }{\tau _d}
\end{equation}
for all ${t,\tau}\in\real_{\ge 0}$ and $t \ge \tau$. \qedp
\end{assumption}

\subsection{Discussion}

The considered assumptions only constrains the attacker action in time 
by posing limitations on the frequency of DoS and its duration. 
Such a characterization can capture many different 
scenarios, including trivial, periodic, random and protocol-aware jamming attacks \cite{A}-\cite{D}. For the sake of simplicity,
we limit out discussion to the case of radio frequency (RF) jammers, although 
similar considerations can be made with respect to spoofing-like threats \cite{Bellardo}.

Consider for instance the case of
\emph{constant jamming}. Constant jamming is one of the most common threats that may occur in a wireless network
\cite{Xu,Pelechrinis}. By continuously emitting RF signals on the wireless medium, 
this type of jammer can lower the Packet Send Ratio (PSR)
for transmitters employing carrier sensing as medium access policy 
as well as lower the Packet Delivery Ratio (PDR) by corrupting packets at the receiver. 
In general, the percentage of packet losses caused by this type of jammer
depends on the Jamming-to-Signal Ratio and can be difficult to quantify 
as it depends, among many things, on the type
of anti-jamming devices, the possibility to adapt the signal strength threshold
for carrier sensing, and the interference signal power, which may vary with time. 
In fact, there are several provisions that can be taken
in order to \emph{mitigate} DoS attacks, including spreading techniques, high-pass filtering
and encoding \cite{C,D}. These provisions 
decrease the chance that a DoS attack will be successful,
and, as such, limit in practice the frequency and duration of the time intervals
over which communication is effectively denied. This is nicely 
captured by the considered assumptions.

As another example, consider the case of \emph{reactive jamming} \cite{Xu,Pelechrinis}.
By exploiting the knowledge of the 802.1i MAC layer protocols, a jammer may  
restrict the RF signal to the packet transmissions. The collision period need not be long since 
with many CRC error checks a single bit error can corrupt an entire frame.
Accordingly, jamming takes the form of a (high-power) burst of noise,
whose duration is determined by the length of the symbols to corrupt \cite{C,Wood}. 
Also this case can be nicely accounted for via the considered assumptions.


\section{Main Results}
\subsection{Modified Consensus Protocol}
The consensus protocol in \eqref{eq:modelloA-discAA} needs to be 
modified in order to achieve robustness against DoS. In this respect, for every $t$ such that the set $\S(\theta,t)=\setdef{i\in I}{\theta_i=0}$ is not nonempty, the nominal 
discrete evolution is modified as follows:
\be\label{eq:modelloA-disc}
\begin{cases} x_i(t^+)=x_i(t) \quad \forall i\in I \\
u_i(t^+)=
\begin{cases}
\sign_{\eps}\!\left(\ave_i(t)\right) \quad \text{if}\: i\in \S(\theta,t) \wedge t \in \Theta(0,t)\\
0 \qquad \qquad \: \qquad \,\text{if}\: i\in \S(\theta,t) \wedge t \in \Xi(0,t)\\
u_i(t)  \qquad \qquad \quad  \text{otherwise}
\end{cases}\\
 \theta_i(t^+)=\begin{cases} f_i(x(t)) \quad \text{if}\: i\in \S(\theta,t) \wedge t \in \Theta(0,t)\\
\frac{\eps}{4d_{i}} \:\, \,\qquad \text{ if}\: i\in \S(\theta,t) \wedge t \in \Xi(0,t)\\
\theta_i(t) \quad \quad \text{otherwise}
\end{cases}\\
\end{cases}
\ee

In words, when a network node attempts to communicate 
and communication is denied, the control signal is set to zero 
until the subsequent attempt
\footnote{It is worth noting that this implicitly requires that 
the nodes be able to detect the DoS status. This is the case, 
for instance, when jamming causes the channel to be busy.
Then, transmitters employing carrier sensing as medium access policy
can detect the DoS status. Another example is when transceivers 
employ TCP acknowledgment.}.

It is worth noting that, 
although an absolute time variable $t$ is used in the description of the 
system dynamics, the various nodes implementing the consensus protocol need not to be aware of such an absolute time. Instead, they rely on their local clocks $\theta_i$. As the nodes rely on their local clocks $\theta_i$ the jump times of each variable $\theta_i$ naturally define a sequence of local switching times, which we denote by $\{t^i_k\}_{k\in \integernonnegative}$.
In particular, we have
\be\label{eq:st}
t_{k+1}^i=t_k^i +\left\{ \begin{array}{l}
{f_i}(x(t_k^i))\;\;\,\;\;t_k^i \in \Theta(0,t)\\ \\
\displaystyle \frac{\varepsilon }{{4{d_{i}}}}\;\;\;\;t_k^i \in \Xi(0,t)
\end{array} \right. ~~\forall~i\in I. 
\ee
The modified algorithm basically consists of a two-mode 
sampling logic, where the sampling times in presence of DoS are chosen 
different, possibly much smaller, than in the nominal situation where DoS is absent. 
As it will become clear later on, this is in order to maximize the
robustness of the consensus protocol against DoS.
By (\ref{eq:st}), it is an easy matter to see that 
the sequences of local switching times $\{t^i_k\}_{k\in \integernonnegative}$ 
satisfy a ``dwell time'' property since 
\be\label{eq:DeltaT-bound}
\Delta_{k}^i:=t^i_{k+1}-t^i_{k} \, \ge \, =\frac{\eps}{4 \degmax}.
\ee   
for every $i \in I$ and $k\ge 0$, where $d_{max}=\max_{i \in I} d_i$.

For the sake of clarity, the modified consensus protocol is summarized 
below.

\vspace{0.4cm}

\medskip\myrule
\vspace{.75\smallskipamount}
\noindent\hfill\textbf{Modified Consensus Protocol \quad (for each $i\in I$\,)}
\hfill\vspace{.75\smallskipamount}
\myrule\vspace{.75\smallskipamount}
\begin{algorithmic}[1]
\STATE {\bf initialization:} set
$u_i(0)\in \{-1,0,+1\}$ and $\theta_i(0)=0$;
\WHILE{$\theta_i(t)>0$} 
\STATE $i$ applies the control $u_i(t)$;
\ENDWHILE
\IF{$\theta_i(t)=0$ \& $t\in \Theta(0,t)$}
\FORALL{$j\in \neigh{i}$} 
\STATE $i$ polls $j$ and collects the information $x_j(t)-x_i(t)$;
\ENDFOR
\STATE $i$ computes $\ave_i(t)$; 
\STATE $i$ computes $\theta_i(t^+)$ as in \eqref{eq:modelloA-disc};
\STATE $i$ computes $u_i(t^+)$ as in \eqref{eq:modelloA-disc};
\ELSE \IF {$\theta_i(t)=0$ \& $t\in  \Xi(0,t)$}
\STATE $i$ set $u_i(t^+)=0$;
\STATE $i$ set $\theta_i(t^+)=\frac{\varepsilon}{4d_i}$;
\ENDIF
\ENDIF
\end{algorithmic} 
\vspace{.5\smallskipamount}\myrule\smallskip
\stopmodif
\smallskip

\vspace{0.4cm}

We are now in position to characterize the overall network behavior 
in the presence of DoS. In this respect, the analysis is subdivided 
into two main steps: i) we first prove that 
all the network nodes eventually stop to update their 
local controls; and ii) we then provide conditions on the DoS frequency 
and duration such that consensus, in the sense of \eqref{set.E}, is preserved.
This is achieved by resorting to a notion of
\emph{Persistency-of-Communication} (PoC), which 
stipulates that disruptions of the graph connectivity 
cannot exceed a prescribed threshold. 
For convenience the proofs are reported in
Section III-B.

As for ii), the following result holds true. 

\begin{proposition} \label{thm:modelloA-convergence}
(\emph{Convergence of the solutions})
Let $x(t)$ be the solution to \eqref{eq:modelloA-cont} and \eqref{eq:modelloA-disc}. 
Then, for every initial condition $x(0)$, there exists 
a finite time $T_1$ such that $u_i(t)=0$ for all $t>T_1$ and $i \in I$.
\qedp
\end{proposition} 

Clearly, the above result does not allows one to conclude 
anything about the final disagreement vector in the sense
that given a pair of nodes $(i,j)$ the asymptotic value 
of $x_i-x_j$ can be arbitrarily large. In order to recover 
the same conclusions as in Theorem 1,  
bounds on the parameters $\tau_d$ and $\tau_f$
in Assumption 1 and 2 have to be enforced.

Consider a sequence $\{t_k^i\}_{k \in \integernonnegative}$ of sampling times, along with a DoS sequence $\{h_n\}_{n \in \integernonnegative}$. Let
\begin{eqnarray} \label{sampling_over_DoS}
 {\mathscr S} \, := \,  \left\{ (i,k) \in I \times \integernonnegative \, |\,\,\, t_k^i  \in 
\bigcup_{n \in \mathbb Z_{\ge 0}} H_n \right\}
\end{eqnarray} 
denote the set of integers related to a control update attempt occurring under DoS. 
Notice that, due to the finite sampling rate, a time interval will necessarily elapse from the time $h_n+\tau_n$ at which DoS ceases, to the time at which the nodes successfully sample and transmit. By construction, this interval 
can be upper bounded as
\begin{eqnarray}  \label{DoS_intervals_delay}
 \sup \limits_{(i,k) \in \mathscr S} 
 \Delta_k^i  \, \le \,  \Delta_*
\end{eqnarray}
Notice  now that the last transmission attempt (if any) of node $i$ over the 
$n$-th DoS interval necessarily falls within
$[h_n+\tau_n-\frac{\eps}{4 d_i},h_n+\tau_n[$.
This means that the next sampling falls by construction
within $[h_n+\tau_n,h_n+\tau_n+\frac{\eps}{4 d_i}[$.
Hence, we conclude that a  DoS free interval of a length greater 
than $\Delta_*$ guarantees that all nodes are able to sample and transmit.
Accordingly, define the following sets that account for the DoS-induced actuation delay, 
\begin{eqnarray} \label{eq:Xibar}
\bar \Xi (\tau ,t) :=\mathop  \bigcup \limits_{n\in{\integernonnegative}} 
\bar H_n \bigcap {[\tau,t ]} \\ 
\bar \Theta (\tau ,t) := [\tau,t ] \;  \backslash \; \bar \Xi (\tau ,t) 
\end{eqnarray}
where 
\[
\bar H_n := [h_n,h_n+\tau_n+ \Delta_*[
\]
By the previous arguments, a sufficient condition under which all the network nodes 
are able to communicate at least once within $[\tau,t[$
is that $\bar \Theta (\tau ,t)$ has positive measure.

The following result then holds.

\begin{proposition} \label{prop:persist} 
(\emph{Persistency-of-Communication})
Let $x(t)$ be the solution to \eqref{eq:modelloA-cont} and \eqref{eq:modelloA-disc}.
Consider any DoS sequence satisfying Assumption 1 and 2 
with 
\begin{equation}\label{eq:DoScons}
\phi(\tau_f,\tau_d, \Delta_*) \, := \, 
\frac{1}{\tau_d}+\frac{\Delta_*}{\tau_f}<1
\end{equation}   
and $\eta$ and $\kappa$ arbitrary. Then,
for every $\tau$, the set
$\bar \Theta (\tau ,t)$ has positive measure for any time $t$ satisfying
\begin{equation} \label{eq:Consensus-Time}
t>\tau+(\kappa+(1+\eta) \Delta_*)
\left( 1-\phi(\tau_f,\tau_d, \Delta_*) \right)^{-1}
\end{equation} \qedp
\end{proposition}

Combining Proposition 1 and 2, the main result of this paper
follows at once. 

\begin{theorem} 
Let $x(t)$ be the solution to \eqref{eq:modelloA-cont} and \eqref{eq:modelloA-disc}.
Consider any DoS sequence that satisfies Assumption 1 and 2 with 
$\tau_f$ and $\tau_d$ as in (\ref{eq:DoScons})
and $\eta$ and $\kappa$ arbitrary. Then,
for every initial condition, $x(t)$ converges in finite 
time to a point $x^*$ belonging to the set $\mathcal E$ 
as in (\ref{set.E}). \qedp
\end{theorem}

A few remarks are in order.

\begin{remark} 
Condition (\ref{eq:DoScons}) in Proposition 2 amounts to 
requiring that the DoS signal does not destroy communication in a 
persistent way. This requirement is indeed reminiscent of 
\emph{Persistency-of-Excitation} (PoE) conditions that are found in 
the literature on consensus under switching topologies, \emph{e.g.}, \cite{Arcak}.
There are, however, noticeable differences. In the present 
case, the incidence matrix of the graph is a time-varying 
matrix satisfying: i) $D(t)=0$ in the presence of DoS; and ii)
$D(t)=D$ in the absence of DoS, where $D$ represents 
the incidence matrix related to the nominal graph configuration.
Consider now a DoS pattern consisting of countable number of singletons, namely $\Xi(0,t)=\bigcup_{n \in \mathbb Z_{\ge 0}} \{h_n\}$, with $\Lambda_n\le {\Delta_*}$. It is trivial to conclude that there exist 
constant $\delta \in \real_{> 0}$ and $\alpha \in \real_{> 0}$
such that (\emph{cf}.  \cite{Arcak})
\[
\int_{t_0}^{t_0+\delta} Q D(t)D^\top(t) Q^\top= Q DD^\top Q^\top \delta>\alpha I
\]
for all $t_0 \in \real_{\ge 0}$, where $Q$ is a suitable projection matrix. 
However, in accordance 
with the previous discussion, consensus can be destroyed.
The subtle, yet important, difference is due to the constraint 
on the frequency of the information exchange that is 
imposed by the network.
In this sense, the notion of PoC naturally extends the PoE condition 
to digital networked settings by requiring that the graph connectivity
be established over periods of time that are consistent with the 
constraints imposed by the communication medium. \qedp
\end{remark}


\subsection{Convergence Analysis}\label{sect:analysis}

This section is devoted to the proof of Proposition 1,
Proposition 2 and Theorem 2. 

{\em Proof of Proposition 1}.
Let
\[
V(x(t))=\frac{1}{2}{x^T(t) L x(t)}
\]
where $t \ge 0$. Consider the evolution of $\dot V(t)$ along 
the solutions to (\ref{eq:modelloA-cont}). 
Following the same steps as in \cite{Claudio-Paolo}, it is easy to verify that
\be\label{eq:dotV-bound}
\dot V(x(t))\le -\dst\sum_{i:|\ave_{i} (t_k^i)|\ge \eps \; \wedge \; t_k^i \in  \Theta(0,t)} \frac{\eps}{2}
\ee
In words, the derivative of $V$ decreases whenever, for some node $i$, 
two conditions are met: i) $|\ave_{i} (t_k^i)| \geq \eps$, which means 
that node $i$ has not reached the consensus set; and ii) 
communication is possible.

From (\ref{eq:dotV-bound}) we deduce that there must exist 
a finite time $T_1$ such that, for every node $i$ and every $k$ 
with $t_k^i \geq T_1$, either $\lvert \ave_i(t_k^i)\rvert < \varepsilon$ or $t_k^i \in \Xi(0,t)$.
This is because, otherwise, the function $V$ would become negative 
contradicting the fact that $V$ is non-negative definite since $L$ is
the Laplacian graph. Thus the proof follows simply by recalling 
that in both the cases $\lvert \ave_i(t_k^i)\rvert < \varepsilon$ and $t_k^i \in \Xi(0,t)$
the control $u_i$ is set to zero. \qedp

{\em Proof of Proposition 2}
By definition of $\bar \Xi$ and in view of Assumption 1 and 2, 
the following bounds on $\bar \Xi$ is readily obtained:
{\setlength\arraycolsep{0pt} 
\begin{eqnarray} 
|\bar \Xi(\tau,t)| \, &\leq& \, |\Xi(\tau,t)| + (n(\tau,t)+1) \Delta_* \nonumber \\
&\leq& \, \kappa+\frac{t-\tau}{\tau_d} + \left(\eta+\frac{t-\tau}{\tau_f}+1\right) \Delta_*
\end{eqnarray}}%
Finally notice that 
\be \label{eq:com}
\lvert  \bar \Theta(\tau,t) \rvert=t-\tau-\lvert  \bar \Xi(\tau,t) \rvert
\ee
Combining the two equations above, one sees
that a sufficient condition for PoC is that 
$
t - \tau > \lvert  \bar \Xi(\tau,t) \rvert 
$,
which, in turn, is implied by
\be \label{eq:com}
t-\tau > \kappa+\frac{t-\tau}{\tau_d} + \left(\eta+\frac{t-\tau}{\tau_f}+1\right) \Delta_*
\ee
This is equivalent to
\be \label{eq:com2}
\left( 1- \phi(\tau_f,\tau_d,\Delta_* \right) (t-\tau)  > 
\kappa+(1+\eta)\Delta_* 
\ee
which concludes the proof. \qedp 

\emph{Proof of Theorem 2}. 
The proof follows immediately by combining 
Proposition 1 and 2. In fact, by Proposition 1, 
all the local controls converge to zero in a finite time.
In turn, Proposition 2 excludes the possibility 
that this is due to a persistence of the DoS status. 
This means that converge to the set $\mathcal E$ 
is necessarily achieved. \qedp

\section{Numerical Example}
In what follows we see a numerical example of the proposed consensus protocol in presence of DoS. A sustained DoS attack with variable period and duty cycle, generated randomly. The resulting DoS signal has an average duty cycle of $55\%$. 

\begin{figure}
	\centering
		\includegraphics[width=1\columnwidth]{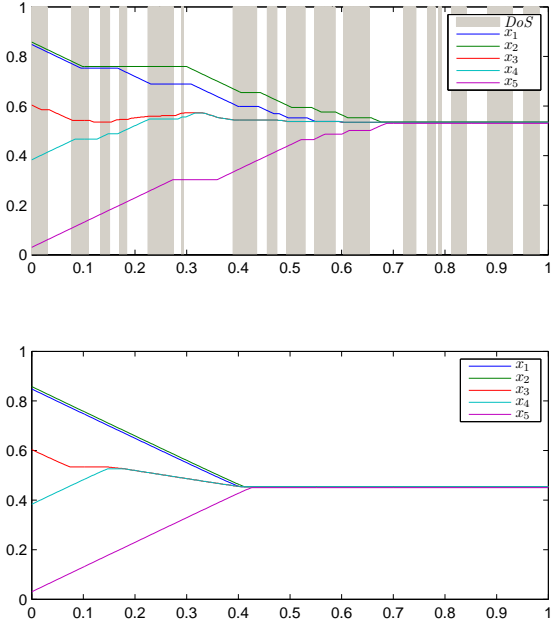}
 \caption{Evolution of state $x$ in \eqref{eq:modelloA-cont} and \eqref{eq:modelloA-disc} with $\varepsilon=0.02$ (a complete graph with n=5 nodes) in presence of DoS with an average duty cycle of$~\sim 55\%$. The
vertical grey stripes represent the time-intervals over which DoS is active.}
	\label{fig:fig1}
\end{figure}

\begin{figure}
	\centering
		\includegraphics[width=1\columnwidth]{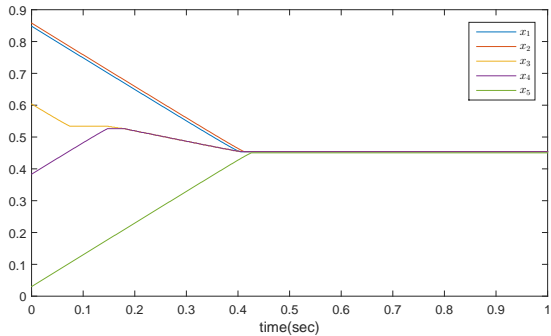}
	\caption{Evolution of state $x$ in \eqref{eq:modelloA-cont} and \eqref{eq:modelloA-disc} with $\varepsilon=0.02$ (a complete graph with n=5 nodes) in absence of DoS.}
	\label{fig:fig2}
\end{figure}

We assume completely connected undirected graph with $n=5$ nodes. During times over which communication is possible each agent is connected to the other agents, namely $d_i=n-1$, while in presence of DoS graph becomes edgeless. 
A sample evolution of \eqref{eq:modelloA-cont} and \eqref{eq:modelloA-disc} starting from the same initial condition and on the same graph is depicted in Fig. \ref{fig:fig1} and Fig. \ref{fig:fig2}. Initial conditions are generated randomly between $0$ and $1$. The vertical gray stripes in Fig. \ref{fig:fig1} represent the time-intervals over which DoS is active.

Consistent with the results in \cite{Claudio-Paolo, Cortes}, system \eqref{eq:modelloA-cont},\eqref{eq:modelloA-disc} converges in finite time to values close to average-max–min-consensus, namely $\frac{1}{2}(\min_i x_i(0) + \max_i x_i(0))$. Presence of DoS bring about latency in coordination of the agents, this is due to controls remaining constantly to zero during this period. Consensus time in Fig. \ref{fig:fig1} is almost twice the consensus time in Fig. \ref{fig:fig2}.

\section{Conclusion}
We investigated coordination of distributed networked systems in the presence of DoS attacks. We argued persistency of excitation condition is not enough to achieve consensus. An explicit characterization of the frequency and duration of DoS attacks under prsistency-of-communication is found. Condition under which agents can transmit information and update their control value frequently enough in sequence of time.   

As an additional future research topic, we can compute the required consensus time taking into account DoS attack. Beside the time cost in \cite{Claudio-Paolo}, this clarifies the consensus time gap in presence and absence of DoS. Furthermore, partial communication failure can also be investigated separately. This problem is motivated by using a point to point communication medium. 



\end{document}